\documentclass[a4,12pt,epsfig]{article}
\usepackage{axodraw,graphicx,psfrag,amssymb,cite}

\renewcommand{\thefootnote}{\fnsymbol{footnote}}

\newcommand{\prepr}[1] {\begin{flushright}  {\bf #1} \end{flushright} \vskip 1.cm}
\newcommand{\titul}[1] {\boldmath \begin{center}{\Large {\bf #1 } } \end{center}
\vskip 0.8cm}

\newcommand{\autor}[1] {\begin{center}  {\bf \lineskip .3cm #1  }
                        \end{center} }

\newcommand{\lugar}[1] {\begin{center}  {\normalsize \bf \it #1   } \end{center}}
\newcounter{muni}

\makeatletter
\def\fmslash{\@ifnextchar[{\fmsl@sh}{\fmsl@sh[0mu]}}
\def\fmsl@sh[#1]#2{%
  \mathchoice
    {\@fmsl@sh\displaystyle{#1}{#2}}%
    {\@fmsl@sh\textstyle{#1}{#2}}%
    {\@fmsl@sh\scriptstyle{#1}{#2}}%
    {\@fmsl@sh\scriptscriptstyle{#1}{#2}}}
\def\@fmsl@sh#1#2#3{\m@th\ooalign{$\hfil#1\mkern#2/\hfil$\crcr$#1#3$}}
\makeatother
\begin{document}
\hbadness=10000
\pagenumbering{arabic}
\begin{titlepage}

\prepr{hep-ph/0308260\\
\hspace{30mm} KIAS--P03049 \\
\hspace{30mm} August 2003}

\begin{center}
\titul{\bf Effect of $H^\pm$ on $D^\pm_s\to \mu^\pm\nu_{\mu}$ and 
$D^\pm_s\to \tau^\pm\nu_{\tau}$
}

\autor{A.G. Akeroyd\footnote{akeroyd@kias.re.kr }}
\lugar{Korea Institute for Advanced Study,
207-43 Cheongryangri 2-dong,\\ Dongdaemun-gu,
Seoul 130-722, Republic of Korea}

\end{center}

\vskip2.0cm

\begin{abstract}
\noindent{We investigate the effect of a charged Higgs boson ($H^\pm$) on the 
decays $D^\pm_s\to \mu^\pm\nu_{\mu}$ and 
$D^\pm_s\to \tau^\pm\nu_{\tau}$, which 
will be measured with high precision at 
forthcoming CLEO-c. We show that a $H^\pm$ can suppress the branching
ratios by $10\%\to 15\%$ from the Standard Model prediction, and 
we emphasize that such contributions should
not be overlooked when comparing lattice calculations 
of $f_{D_s}$ to the values obtained from these decays.
}

\end{abstract}

\vskip1.0cm
\vskip1.0cm
{\bf Keywords : \small Rare D decay} 
\end{titlepage}
\thispagestyle{empty}
\newpage

\pagestyle{plain}
\renewcommand{\thefootnote}{\arabic{footnote} }
\setcounter{footnote}{0}

\section{Introduction}
Purely leptonic decays are the classic ways to measure the
decay constants ($f_P$) of charged pseudoscalar mesons $P^\pm$.
For the light mesons, $\pi^\pm$ and $K^\pm$,
the muonic decays $\pi^\pm\to \mu^\pm\nu_{\mu}$, 
$K^\pm\to \mu^\pm\nu_{\mu}$
have large branching ratios (BRs) and so
their respective decay constants have been determined 
with high precision \cite{Hagiwara:fs} ($< 1\%$). 
For the charmed pseudoscalar mesons ($D^\pm,D^\pm_s$) the 
BRs for the purely leptonic channels
are much smaller than those for the above light mesons 
due to the dominance of weak decay mechanism $c\to W^\pm q$ with
a spectator quark.
These smaller leptonic BRs together with the lack of a dedicated charm
factory has resulted in vastly inferior experimental 
precision for the charmed meson
decay constants compared to that for $f_{\pi}$ and $f_K$.
Current measurements of $f_D$ and $f_{D_s}$
have large errors of around $100\%$ and $15\%$ respectively
\cite{Hagiwara:fs}. 
With the imminent (summer 2003) commencement of the 
CLEO-c experiment \cite{Shipsey:2002ye} 
this situation will
improve dramatically in the next $2\to 3$ years. Precise 
${\cal O} (1\to 2\%)$ measurements of $f_D$ and $f_{D_s}$ 
are expected and will constitute a vital test of lattice methods for the
heavy quark systems, as well as providing crucial experimental
input for calculations of the $B$ meson decay constants 
\cite{Shipsey:2002ye}.

However, absent in the above discussion is the fact that 
the leptonic decays of $D^\pm$ and $D^\pm_s$ 
might be affected by physics beyond the Standard Model (SM).
It is known that new charged particles which couple to the fermions would
contribute at tree--level to these decays\cite{Hou:1992sy}. 
One such example is a charged Higgs boson $H^\pm$, 
and in this paper we 
consider its effect on the decays $D^\pm_s\to 
\mu^\pm\nu_{\mu}$, $D^\pm_s\to\tau^\pm\nu_{\tau}$, and thus the
measured value of $f_{D_s}$. 
We point out that the possibility of such new physics contributions
to these decays should not be overlooked when comparing 
the experimentally measured
value of $f_{D_s}$ to the lattice QCD predictions.

\section{The decays $D^\pm_s\to \mu^\pm\nu_{\mu}$ and
$D^\pm_s \to\tau^\pm\nu_{\tau}$}
Singly charged Higgs bosons, $H^\pm$,
arise in any extension of the SM which
contains at least two $SU(2)\times U(1)$ Higgs doublets, e.g. 
any Supersymmetric (SUSY) model. Together with $W^\pm$ they mediate the 
leptonic decays $D^\pm_{(s)}\to l^\pm\nu_l$ via the
annihilation process shown below:
\begin{center}
\vspace{-50pt} \hfill \\
\begin{picture}(120,90)(0,25) 
\Photon(10,25)(68,25){4}{8}
\Vertex(10,25){3}
\ArrowLine(10,25)(-40,55)
\ArrowLine(-40,-5)(10,25)
\ArrowLine(118,-5)(68,25)
\ArrowLine(68,25)(118,55)
\Text(-18,50)[]{$c$}
\Text(-13,0)[]{$d,s$}
\Text(138,-2)[]{$\tau^\pm,\mu^\pm$}
\Text(129,55)[]{$\nu$}
\Text(41,38)[]{$W^*,H^\pm$}
\GOval(-45,25)(33,10)(0){0.5}
\Text(-70,25)[]{$D^\pm_{(s)}$}
\end{picture}
\end{center}

\vspace*{1.3cm}

The tree--level partial width is given by \cite{Hou:1992sy}: 
\begin{equation}
\Gamma(D^\pm_{(s)}\to \l^\pm\nu_{l})={(G_F^2/8\pi)} m_{D_{(s)}} 
m_l^2 f_{D_{(s)}}^2 r_{(s)}|V_{cd(cs)}|^2 
\left(1-{m_l^2/m^2_{D_{(s)}}}\right)^2
\end{equation}
where $m_l$ is the mass of the lepton, $m_{D_{(s)}}$ is the mass of the
$D^\pm_{(s)}$ meson, $V_{cd(cs)}$ are CKM matrix elements, and
\begin{equation}
r_{(s)}=[1-\tan^2\beta(m^2_{D_q}/m^2_{H^\pm})(m_q/m_c)]^2=
[1-R^2m^2_{D_q}(m_q/m_c)]^2
\end{equation}
where $r_{(s)}=1$ in the SM, $R=\tan\beta/m_{H^\pm}$ and
$\tan\beta=v_2/v_1$ (ratio of vacuum expectation values). 
The $H^\pm$ contribution interferes destructively with 
that of $W^\pm$, causing a suppression in the BR, with
the largest deviations arising for large $R$. For $D^\pm$ this
effect is essentially negligible ($r\approx 1$)
due to the smallness of
$m_d/m_c$, but for $D^\pm_s$ the scaling factor $r_s$ may differ 
from 1 due to the
non--negligible $m_s/m_c$. This has been noted before \cite{Hou:1992sy},
\cite{Hewett:1995aw} but a numerical study was absent.
In light of the high precision expected in the measurement of
these leptonic decays at CLEO-c, we wish to quantify the 
previous qualitative analyses 
in order to see if the $H^\pm$ contribution can
be significantly larger than the anticipated error in the measurement of
BR($D^\pm_s\to \tau^\pm\nu_{\tau},\mu^\pm\nu_{\mu}$). 

The current experimental measurements
and the SM predictions for the three leptonic decays which
CLEO-c expects to measure are given in Table 1. 
For the SM predictions we take the lattice results 
$f_D=226\pm 15$ MeV and  $f_{D_s}=250\pm 30$ MeV
\cite{Ryan:2001ej}, which induces an error of around $15\%\to 25\%$
the BRs. The measurements of the $D^\pm_s$ decays are world averages taken
from Ref.~\cite{Soldner-Rembold:2001zk} and that for 
$D^\pm\to \mu^\pm\nu_{\mu}$
is taken from Ref.~\cite{Bai:cg}. The expected errors from CLEO-c are 
shown in the final column.

\begin{table}
\begin{center}
\begin{tabular} {|c|c|c|c|c|} \hline
Decay & SM BR& Current Exp BR & Exp Error  & CLEO-c Error  \\ \hline
 $D^\pm\to \mu^\pm\nu_{\mu}$ & $4.5\pm 0.6\times 10^{-4}$
& $8^{+16+5}_{-5-2}\times 10^{-4}$ & $\sim 100\%$
 & 3.8\% \\ \hline
 $D^\pm_s\to \mu^\pm\nu_{\mu}$ & $5.2\pm 1.2 \times 10^{-3}$ 
& $5.3\pm 0.9\pm 1.2\times 10^{-3}$ & $25\%$
& 3.2\% \\ \hline
 $D^\pm_s\to \tau^\pm\nu_{\tau}$ & $5.1\pm 1.2\times 10^{-2}$
& $6.1\pm 1.0\pm 0.2 \times 1.3\times 10^{-2}$ & $25\%$
& 2.4\% \\ \hline
\end{tabular}
\end{center}
\caption{SM predictions, current experimental BR, experimental
error and CLEO-c expected errors for certain leptonic decays of 
$D^\pm$ and $D^\pm_{s}$}
\label{explimits}
\end{table}

\section{Numerical Results}
We now quantify the effect of the $H^\pm$ contribution on $r_s$ (eq.2). 
For the quark masses $m_s$ and $m_c$ we use the Particle Data Group values 
\cite{Hagiwara:fs} and obtain $0.06 < m_s/m_c < 0.15$. 
The value of $R (=\tan\beta/m_{H^\pm}$) is best constrained from
non--observation of the decay $B^\pm\to \tau^\pm\nu_{\tau}$,  
giving $R< 0.34\pm 0.02\pm 0.06 \,GeV^{-1}$, where
the first error is from $f_B$ and the second is from 
possible large SUSY corrections \cite{Akeroyd:2003zr}. Thus we take 
$R=0.4\,\,GeV^{-1}$ as our largest value.  

\begin{figure}
\centerline{\includegraphics[width=9cm,height=9cm]{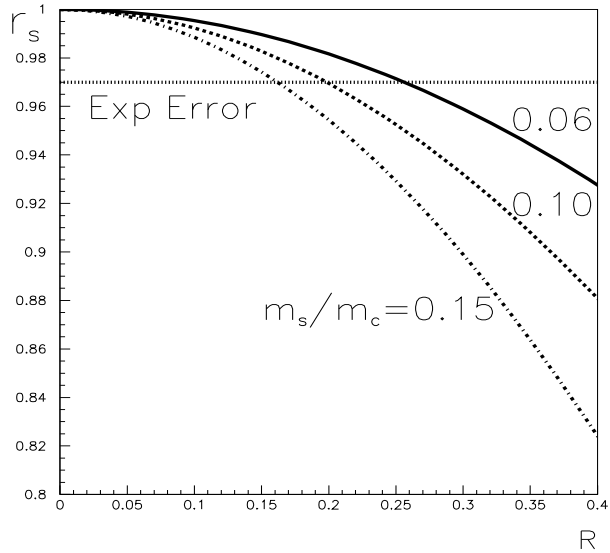}}
\caption{$r_s$ as a function of $R(=\tan\beta/m_{H^\pm})$, 
for various values of $m_s/m_c$}
\label{fig1}
\end{figure} 

In Fig.1 we plot $r_s$ as a function of $R$
for various values of $m_s/m_c$. For $R=0.4$ one has $r_s=0.83 (0.93)$ 
for the largest (smallest) values of $m_s/m_c$. 
This suppression is comfortably larger than the 
anticipated experimental error of $2\%\to 3\%$ (shown by the horizontal
line) in the measurement of 
BR$(D^\pm_s\to \tau^\pm\nu_{\tau},\mu^\pm\nu_{\mu})$.
Thus the presence of $H^\pm$ would lead to a deceptive smaller 
{\sl measured} value of the decay constant $f_{D_s}$.
This effect was pointed out for the case of $f_K$ 
in Ref~.\cite{Hou:1992sy}, where BR($K^\pm\to \mu^\pm\nu_{\mu}$)
can be suppressed by a factor comparable to that for the $m_s/m_c=0.06$ curve. 
Although the effect of $H^\pm$ is less than the $25\%$ error in
BR($D^\pm_s\to \mu^\pm\nu_{\mu},\tau^\pm\nu_{\tau}$) from the
current lattice predictions of $f_{D_s} $\cite{Ryan:2001ej}, 
there are already signs that
the error in $f_{D_s}$ will be significantly improved in the near 
future. A recent paper \cite{Juttner:2003ns} calculated $f_{D_s}$ 
with a precision of $4\%$ $(252\pm 9$ GeV) in the quenched approximation,
while the techniques discussed in Ref.~\cite{Davies:2003ik} 
promise comparable or smaller errors in the unquenched approximation.
With these anticipated reductions in the theoretical 
error of $f_{D_s}$, we suggest that the possible 
effects of any $H^\pm$ should not be overlooked when comparing the 
experimentally extracted $f_{D_s}$ to the prediction from lattice QCD.

An additional observable which will also be a test of lattice QCD is the
ratio of the muonic decay rates ${\cal R}_{\mu}$ defined by
\begin{equation}
{\cal R}_{\mu}={BR(D^\pm_s\to \mu^\pm\nu_\mu)/BR(D^\pm\to \mu^\pm\nu_\mu)}
\sim (f_{D_s}/f_{D})^2
\end{equation}
The lattice prediction for $f_{D_s}/f_{D}$
is known with substantially greater precision than the 
individual values of the decay constants, and currently
stands at $1.12 (4)$ for unquenched calculations and 
$1.12(2)$ in the quenched approximation \cite{Ryan:2001ej}, i.e.
an error $< 4\%$. 
A similar ratio (${\cal R}_{\tau}$) for the decays 
$D^\pm_{(s)}\to \tau^\pm\nu_{\tau}$ is also potentially an 
experimental observable,
but is unlikely to be measured in the foreseeable future since CLEO-c has
limited sensitivity to 
$D^\pm\to \tau^\pm\nu_{\tau}$ \cite{Shipsey:2002ye}. 
Hence we will only consider ${\cal R}_{\mu}$, whose current 
SM prediction is given by ${\cal R}_{\mu}=12\pm 0.8$, i.e. 
$\sim 7\%$ error.
The current experimental measurement of ${\cal R}_\mu$ is based on
1 event for BR$(D^\pm\to \mu^\pm\nu_\mu)$ \cite{Bai:cg}, 
whose central value is consistent with the old MARKIII
limit of BR$(D^\pm\to \mu^\pm\nu_{\mu})< 7.2\times 10^{-4}$ 
\cite{Adler:1987ty}.
Using the latter, a current lower bound would be ${\cal R}_\mu > 7\pm 2$.
The first accurate measurement of $R_{\mu}$ is expected 
at CLEO-c with an error of around $7\%$, which is roughly the same as
the error in the lattice prediction for ${\cal R_{\mu}}$.
In contrast, in the case of the individual
BRs the current theoretical error is substantially larger than the 
expected experimental error. The presence of $H^\pm$ would modify 
${\cal R}_{\mu}$ by the factor $r_s$.
Since the expected theoretical error in ${\cal R}_{\mu}$ 
should approach the percent level or less, ${\cal R}_{\mu}$ 
may also be a sensitive probe of physics beyond the SM.
As an example, in SUSY models with $R$ Parity violating slepton interactions,
BR($D^\pm\to \mu^\pm\nu_{\mu})$, which is essentially unaffected
by $H^\pm$, can be significantly 
suppressed or enhanced by the coupling combination 
$\lambda_{232}\lambda'_{221}$,
as discussed in Ref~.\cite{Akeroyd:2002pi}. 
Thus the presence of these couplings would give rise 
to a larger ${\cal R_\mu}$ ($>> 12$)
or allow values close to the current experimental limit 
$({\cal R}_\mu \approx 7)$ depending on the sign and magnitude 
of the product of $R$ Parity violating couplings $\lambda\lambda'$. 
Thus the first measurements of $R_{\mu}$ from 
CLEO-c are eagerly awaited.

Finally we note that any sizeable effects of 
$H^\pm$ on BR$(D_s^\pm\to \mu^\pm\nu_{\mu},\tau^\pm\nu_{\tau}$) and
${\cal R}_{\mu}$ should manifest themselves in the purely
leptonic $B^\pm$ decays, $B^\pm\to \tau^\pm\nu_{\tau},\mu^\pm\nu_{\mu}$.
This is because $r_s$ depends strongly on $R (=\tan\beta/m_{H^\pm}$),
whose permitted value is constrained from the upper limits on the
above $B^\pm$ decays. The $B$ factories will be sensitive to 
$R\sim 0.25$ with 400 fb$^{-1}$, and thus any significant suppression 
in $r_s$ from $H^\pm$ would be accompanied by a corresponding
enhancement in $B^\pm\to \tau^\pm\nu_\tau$ (and $\mu^\pm\nu_\mu$).

\section{Conclusions}
We have studied the effect of a $H^\pm$ on the leptonic decays
$D^\pm_{s}\to \mu^\pm\nu_\mu,\tau^\pm\nu_{\tau}$. 
We showed that $H^\pm$ can suppress the BRs by up to $10\%\to 15\%$,
which is larger than the expected experimental error ($2\%\to 3\%$) from 
CLEO-c. We suggested that new physics effects like these 
should not be overlooked when
comparing the experimental measurements of $f_{D_s}$
to the SM lattice QCD predictions.


\end{document}